\documentclass[conference]{IEEEtran}
\IEEEoverridecommandlockouts
\usepackage{cite}
\usepackage{amsmath,amssymb,amsfonts}
\usepackage{algorithmic}
\usepackage{graphicx}
\usepackage{textcomp}
\usepackage{xcolor}
\usepackage{amsmath,lipsum}
\usepackage[ruled]{algorithm2e}
\usepackage{graphicx}
\usepackage{float}
\usepackage{subfigure}
\usepackage{multirow}
\usepackage{multicol}
\usepackage{arydshln}
\usepackage{cite}
\usepackage{stfloats}

\def\BibTeX{{\rm B\kern-.05em{\sc i\kern-.025em b}\kern-.08em
    T\kern-.1667em\lower.7ex\hbox{E}\kern-.125emX}}
\begin{document}

\title{Receptive Field-based Segmentation for Distributed CNN Inference Acceleration in Collaborative Edge Computing\\ }
\author{\IEEEauthorblockN{Nan~Li, Alexandros~Iosifidis and Qi~Zhang }
\IEEEauthorblockA{DIGIT, Department of Electrical and Computer Engineering, Aarhus University.\\
Email: \{linan, ai, qz\}@ece.au.dk}}
\maketitle

\begin{abstract}
This paper studies inference acceleration using distributed convolutional neural networks (CNNs) in collaborative edge computing network. To avoid inference accuracy loss in inference task partitioning, we
propose receptive field-based segmentation (RFS). To reduce the computation time and communication overhead, we propose a novel collaborative edge computing using fused-layer parallelization to partition a CNN model into multiple blocks of convolutional layers. In this scheme, the collaborative edge servers (ESs) only need to exchange small fraction of the sub-outputs after computing each fused block. In addition, to find the optimal solution of partitioning a CNN model into multiple blocks, we use dynamic programming, named as dynamic programming for fused-layer parallelization (DPFP). The experimental results show that DPFP can accelerate inference of VGG-16 up to 73\% compared with the pre-trained model, which outperforms the existing work MoDNN in all tested scenarios. Moreover, we evaluate the service reliability of DPFP under time-variant channel, which shows that DPFP is an effective solution to ensure high service reliability with strict service deadline.
\end{abstract}


\section{Introduction}
Convolutional neural networks (CNNs) have been widely studied in performing emerging Internet of Things (IoT) applications, such as autonomous driving and augmented reality. However, running computationally intensive CNNs solely at resource-constrained IoT devices is often infeasible to meet stringent deadline in time-critical applications \cite{005}. 
To solve this issue, various neural network model compression techniques were proposed to simplify CNN topologies and reduce computing operations \cite{006}. However, loss of accuracy is inevitable.

Edge computing is a promising approach that the computation intensive tasks are offloaded to edge server (ES) to reduce inference time \cite{008,Qi2015IoT}. However, the offloading channel state is stochastic in practice, which may cause fluctuations in offloading time and consequently result in missing the service deadline \cite{Nan2017CSI,jianhui2020Access}. Accelerating CNN inference is a viable approach to address the uncertainties in offloading time, thereby ensuring high service reliability in time-critical IoT applications. 
\begin{table}[t]
\centering
	\caption{Inference accuracy of VGG-16 using different Distributed CNN methods for two ESs with equal computing capacity. }
	\label{accuracy}
	\scalebox{0.9}{
	\begin{tabular}{|c|c|c|}
		\hline
		Method & Top-1 Acc. &Top-5 Acc. \\
		\hline
		Pre-trained model & \textbf{71.5\%} & \textbf{92.7\%} \\
		\hline
		Kernel-size based segmentation \cite{015,030}  & 53.2\% & 67.9\% \\
		\hline
		Computing-power based segmentation \cite{005,016} & 60.1\%  & 78.4\% \\
		\hline
		{\textbf{Proposed RFS}} & \textbf{71.5\% } & \textbf{92.7\%} \\
		\hline
	\end{tabular}}
\vspace{-5mm}
\end{table}

This paper proposes collaborative edge computing for distributed CNN inference acceleration by leveraging the fundamental processing characteristics of CNN. For a convolutional layer (CL), the input is a tensor with the shape (height, width, channel). Note that the input of the first CL is the original image with 3 channels (R, G, B). The convolution operation of a CL is basically a {\textit{dot product}} between the kernel-sized patch of the input and the kernel, which is then summed, always resulting in one entry of the output tensor. Applying the kernel multiple times across the spatial dimensions (width and height) of the input tensor, a 2-dimensional feature map of the output tensor for each kernel can be generated. Theoretically speaking, the computation of a CL can be partitioned and processed by different computation units collaboratively without loss of inference accuracy. Now the question is how to partition an inference task and distribute it among the ESs, so that it can minimize the overall inference time while keeping the same inference accuracy as the pre-trained model.

To partition a CNN inference task, segment-based spatial partitioning \cite{015,016} can be used to partitions an input tensor along the largest dimension, e.g., height or width. Two examples of segment-based spatial partitioning, kernel-size based segmentation \cite{015,030} and computing-power based segmentation \cite{005,016}, are proposed to partition an input tensor flexibly depending on the computation resource of each ES and the number of available ESs. We implemented the existing works and tested their inference accuracy, as shown in Table \ref{accuracy}. It can be seen that the existing works ignore the effect of stride and padding for each CL which can cause intolerable inference accuracy loss. To solve this issue, we propose a receptive field-based \cite{031} segmentation (RFS) which achieves the exact same result as the pre-trained model, namely without scarifying any inference accuracy. Note that all methods following the approach used in \cite{005,015,016,030} are expected to suffer from similar performance drops. For this reason, hereafter we consider MoDNN as a representative methods of this category of methods and conduct extensive comparisons with our proposed RFS which achieves exact calculation of the model's output.

\begin{figure*}[htbp]
\centering
\includegraphics[width=0.75\textwidth]{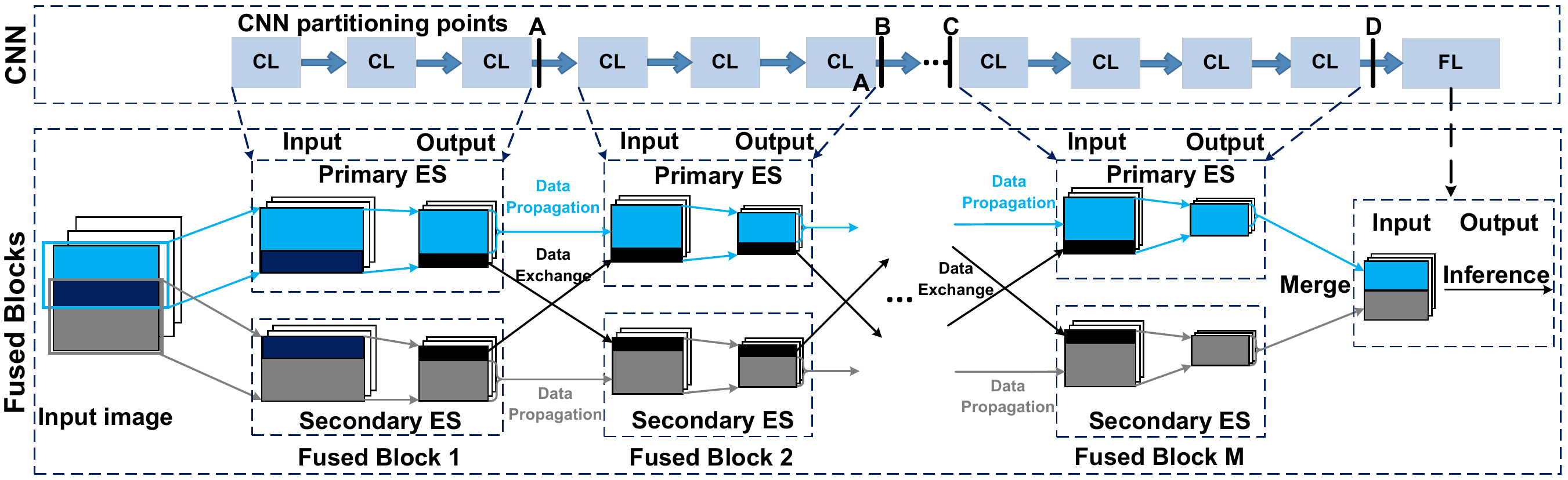}
\caption{Collaborative edge computing based on fused-layer parallelization}
\label{Fig.1}
\vspace{-4mm}
\end{figure*}

To parallelize the sub-tasks among ESs, the fused-layer parallelization (i.e., several CLs are fused into one block, referred to as \textit{fused block}) was used in the previous works \cite{030}. However, in these works the sub-output tensors of all secondary ESs are entirely sent back to primary ES and merged into a new input tensor, then the primary ES partitions the new input tensor into several sub-input tensors and distributes each of them to individual ES. This will result in substantial communication overhead. To further reduce the communication cost, we propose the collaborative edge computing for fused-layer parallelization (in Fig. \ref{Fig.1}), in which only a small fraction of the sub-output on each ES needs to be exchanged with the other collaborating ESs after computing each fused block. This is because each ES already has most of its input to proceed with the next fused block; therefore, it only needs to obtain the missing information that is needed to proceed with the next fused block from its collaborating ESs.

This paper studies inference acceleration using distributed CNNs in collaborative edge computing network. Our contributions are summarized as follows,
\begin{itemize}
	\item We propose RFS to partition a CNN inference task among the collaborative ESs without compromising the inference accuracy of the pre-trained model.
	\item We propose a novel fused-layer parallelization using collaborative edge computing which reduces the computation time and communication overhead. We formulate an optimization problem to minimize the overall inference time and use dynamic programming for fused-layer parallelization (DPFP) to solve this problem.
	\item We conduct experiments on the high-end GPUs RTX 2080TI and GTX 1080TI, as well as on the embedded GPU JETSON AGX Xavier, to measure the computation time. The experimental results show that our solution can accelerate inference of VGG-16 up to 73\% compared with a standalone ES running the pre-trained models. This outperforms the state-of-the-art work MoDNN \cite{005}.
	\item We evaluate the service reliability of DPFP under time-variant channel. It shows that DPFP is effective to ensure high service reliability with strict service deadline.
\end{itemize}

The remainder of this paper is organized as follows. Section \ref{systemmodel} presents the system model of distributed CNN inference using collaborative ESs. In Section \ref{problem}, we propose a novel fused-layer parallelization using collaborative edge computing and formulate the optimizing problem, then in Section \ref{solution} we use DFDP to solve this problem. The simulation results are presented and discussed in Section \ref{simulation}, and the conclusions are drawn in Section \ref{conclusion}. The source code will be made publicly available at https://gitlab.au.dk/netx/agileiot/dpfp.git.
\section{System Model}\label{systemmodel}
This paper studies a collaborative edge computing network, in which an IoT device can offload CNN inference task to a primary ES, which can partition the task and distribute the sub-tasks to multiple secondary ESs and let them collaboratively solve the inference task, as shown in Fig. \ref{Fig.1}. The inference task is assumed to use a CNN ${\bf{G}}$ with $N$ CLs and several fully connected layers (FLs). The connection between the primary ES and the secondary ESs is high-speed Ethernet. The set of ESs can be denoted as $ {\bf{E}} = \left\{ {{e_1}, \cdots ,{e_K}} \right\}$, where ${e_1}$ represents the primary ES and the others are secondary ESs. To accelerate a CNN inference task, the primary ES can partition the task into several sub-tasks, each assigned to and processed in parallel on the selected ES, as shown in Fig. \ref{Fig.1}.
\subsection{How to parallelize CNN computation in ESs}
To reduce communication cost after each CL, we use the fused-layer parallelization to partition CNN model {\bf{G}} into $M+1$ fused blocks, where the last fused block is composed of all the FLs ($M <=N$). In our proposed scheme (Fig. \ref{Fig.1}), the primary ES is responsible for deciding the data size to be exchanged. The set of the fused blocks is denoted as ${\bf{F}} = \left\{ {{f_1},{f_2},...,{f_{M+1}}} \right\}$, where ${f_m}$ is the $m{\rm{th}}$ fused block. We use $l_{{f_m}}^{s}$ and $l_{{f_m}}^{e}$ to denote the start layer index and end layer index of fused block ${f_m}, m \neq M+1$, which subject to
\begin{equation}
	l_{f_m}^s = \left\{ {\begin{array}{*{20}{c}}
			1, & f_m = f_1,\\
			 l_{f_{m-1}}^e + 1, & f_m \in {\bf{F}}\setminus\left\{f_1, f_{M+1}\right\}.
	\end{array}} \right.
	\label{eq5}
\end{equation}
\newcounter{TempEqCnt}
\setcounter{TempEqCnt}{\value{equation}}
\setcounter{equation}{12}
\begin{figure*}[hb]
\begin{equation}
	{\textit{EX}}_{{f_m}}^{{e_k},{e_{k - 1}}} = 4\left[ \max \left({\textit{OE}}_{f_{m-1}}^{e_{k - 1}} - {\textit{IS}}_{f_m}^{e_k},0 \right)+1\right] {\textit{IF}}_{f_m} c_{f_m}^{\textit{in}}, \;{f_m} \in {\bf{F}}\setminus \left\{ {f_1},{f_{M+1}}  \right\}.
\end{equation}
\begin{equation}
	{\textit{EX}}_{{f_m}}^{{e_k},{e_{k + 1}}} = 4\left[ \max \left({\textit{IE}}_{f_m}^{e_k} - {\textit{OS}}_{f_{m -1}}^{e_{k + 1}},0 \right)+1\right]  {\textit{IF}}_{f_m} c_{f_m}^{\textit{in}},\; {f_m} \in {\bf{F}}\setminus \left\{ {{f_1},{f_{M+1}}} \right\}.
\end{equation}
\begin{equation}
		{S}\left( {{f_m},{\bf{E}}} \right) = \left\{ {\begin{array}{*{20}{c}}
				{\sum\limits_{{e_k} \in {\bf{E}}\setminus \left\{ {{e_1}} \right\}} 4{\left( {{\textit{IE}}_{f_m}^{{e_k}} - {\textit{IS}}_{f_m}^{{e_k}} + 1} \right) {\textit{IF}}_{f_m}  c_{f_m}^{\textit{in}}} },&{{f_m} = {f_1}},\\
				{\sum\limits_{k \in {\bf{E}}} {\textit{EX}}_{f_m}^{{e_k},{e_{k - 1}}} + {\textit{EX}}_{f_m}^{{e_k},{e_{k + 1}}}}, &{{f_m} \in {\bf{F}}\setminus \left\{ {{f_1},{f_{M+1}}} \right\}},\\
				{\sum\limits_{{e_k} \in {\bf{E}}\setminus \left\{ {{e_1}} \right\}} 4{\left( {{\textit{OE}}_{f_M}^{{e_k}} - {\textit{OS}}_{f_M}^{{e_k}} + 1} \right) {\textit{OF}}_{f_M}  c_{f_M}^{\textit{out}}} },&{{f_m} = {f_{M+1}}}.
		\end{array}} \right.
	\end{equation}
	\vspace{-3mm}
\end{figure*}
\subsection{RFS based CNN task partitioning}
In CNN, every kernel (or filter) looks at a specific part of the input tensor (named as \textit{Receptive field}), performs multiplication-addition operations, and then moves by a defined number of pixels (stride). Assuming the kernel size, the padding size and the stride size of CL $i$ are $k_{i} \times k_{i}$, $p_{i}$ and $s_{i}$, respectively, the attributes of CL $i$ can be calculated by\cite{034}:
\setcounter{equation}{1}
\begin{equation}
{\it{OF}_{i}} = \left\lfloor {\left({\it{IF}_{i}} + 2p_{i} - k_{i}\right)}/{s_{i}} \right\rfloor  + 1,
\label{eq1}
\end{equation}
\begin{equation}
{j_{i}} = {j_{i-1}}  s_i,
\label{eq2}
\end{equation}
\begin{equation}
{r_{i}} = {r_{i-1}} + \left( {k_i - 1} \right) {j_{i-1}},
\label{eq3}
\end{equation}
\begin{equation}
\sigma_{i} = \sigma_{i-1} + \left[ \left({k_i - 1}\right)/{2} - p_{i} \right] {j_{i-1}},
\label{eq4}
\end{equation}
where ${\it{OF}_{i}}$ is both the width and height of the output feature of CL ${i}$, $j_{i}$ is the cumulative stride (referred to as \textit{jump}) in the output tensor, $r_{i}$ is the receptive field size of the output tensor, and $\sigma_{i}$ is both the row index and column index of the center position of the receptive field of the first output feature. Note that the input tensor of CL $i$ is the output tensor of CL $i-1$, that means, ${\textit{IF}_i} = {\textit{OF}_{i-1}}$.

\begin{figure}[h]
	\centering
	\includegraphics[width=0.35\textwidth]{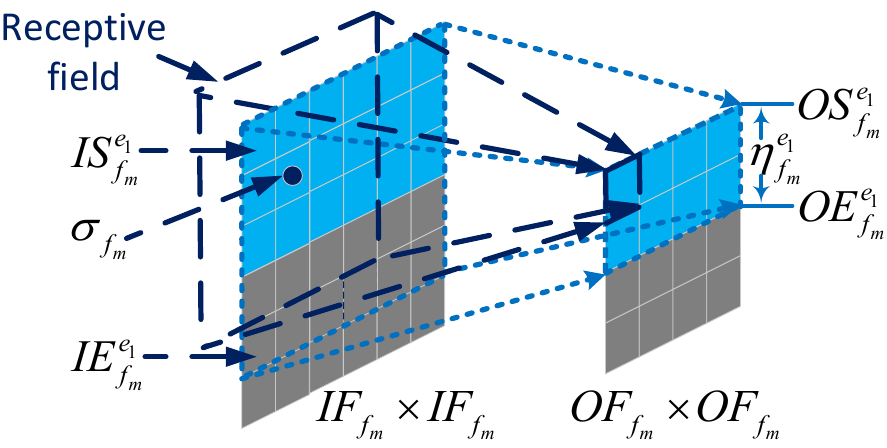}
	\caption{Receptive field-based partitioning for a fused block.}
	\label{Fig.2}
	\vspace{-2mm}
\end{figure}

To partition the task efficiently among ESs, taking into account the communication cost for distributing the task and exchanging the intermediate outputs among the ESs, we propose RFS to partition the input tensor into sub-inputs along the largest dimension of the input. An example of partitioning an input tensor into two sub-inputs is shown in Fig. \ref{Fig.2}.

We define a variable $\eta _{f_m}^{{e_k}}$ to denote the ratio of the sub-output on ES $e_k$ to the entire output of the fused block ${f_m}, m \neq M+1$ which is subject to
\setcounter{equation}{5}
\begin{equation}
	\begin{array}{*{20}{c}}
		{\sum\limits_{{e_k} \in {\bf{E}}} {\eta _{{f_m}}^{{e_k}}}  = 1},\ {\forall {f_m} \in {\bf{F}}\setminus\left\{f_{M+1}\right\}},
	\end{array}
	\label{eq8}
\end{equation}
\begin{equation}
	\begin{array}{*{20}{c}}
		0 \leq \eta _{{f_m}}^{{e_k}}\leq 1 , \ {\forall {f_m} \in {\bf{F}}\setminus\left\{f_{M+1}\right\}} \; {\rm{ and }}
	\end{array} \forall {e_k} \in {\bf{E}}.
	\label{eq9}
\end{equation}

Assuming both the height and width of the output tensor are $\textit{OF}_{{f_m}}$, the start and end row index $\textit{OS}_{{f_m}}^{{e_k}}$ and $\textit{OE}_{{f_m}}^{{e_k}}$ of the sub-output tensor of ES ${e_k}$ can be
computed as
\begin{equation}
	{{\textit{OS}}_{f_m}^{e_k} \hspace{-1.5mm}= \hspace{-1.5mm} \sum\limits_{i = e_1}^{e_{k - 1}} {\eta _{f_m}^i {\textit{OF}}_{f_m}}  \hspace{-1mm}+\hspace{-1mm} 1}, {e_k \in {\bf{E}}\:\:{\rm{ and }}\:\: \hspace{-1mm}{f_m} \in {\bf{F}}\hspace{-1mm}\setminus\hspace{-1mm}\left\{f_{M+1}\right\}}\hspace{-0.5mm},
	\label{eq10}
\end{equation}
\begin{equation}
	{{\textit{OE}}_{f_m}^{e_k} \hspace{-1mm}=\hspace{-1mm} \sum\limits_{i = {e_1}}^{e_k} {\eta _{f_m}^i {\textit{OF}}_{f_m}} },  \:\: {e_k \in {\bf{E}} \:\: {\rm{ and }} \:\: {f_m} \in {\bf{F}}\hspace{-1mm}\setminus\hspace{-1mm}\left\{f_{M+1}\right\}}.
	\label{eq11}
\end{equation}

We assume the receptive field size and the jump of each output pixel for fused block $f_m, f_m \neq f_{M+1}$ are $r_{f_m}$ and ${j_{{f_m}}}$, and the center position of the receptive field of the first output feature is ${\sigma_{{f_m}}}$. The start index and end index of the sub-input on ES $e_k$, $\textit{IS}_{f_m}^{e_k}$ and $\textit{IE}_{f_m}^{e_k}$, can be calculated as follows,
\begin{equation}
	\textit{IS}_{f_m}^{e_k} =  \sigma_{f_m} +\left( \textit{OS}_{f_m}^{e_k}-1 \right) j_{f_m} - \left\lfloor \left(r_{f_m}-1\right)/2 \right\rfloor
\end{equation}
\begin{equation}
\textit{IE}_{f_m}^{e_k} =  \sigma_{f_m}  +\left( \textit{OE}_{f_m}^{e_k}-1 \right) j_{f_m} + \left\lfloor \left(r_{f_m}-1\right)/2 \right\rfloor
\end{equation}

\section{Problem Formulation}\label{problem}
To effectively minimize the overall inference time of a CNN inference task, the key is to trade off the communication and computation time used in processing CLs. This is because increasing number of
ESs can in general reduce the computation time; however, it can result in more communication overhead due to more exchanged data among ESs.
Similarly, increasing the number of fused blocks can reduce computation time while increasing exchanged data size consequently increasing communication time. This section firstly calculates the exchanged data size between ESs, then formulates the joint optimization problem to minimize the overall inference time.
\subsection{Exchanged data size between ESs}\label{section2}
For a tensor $\left(\textit{height}, \textit{width},\textit{channel} \right)$ with type in float32, its size is calculated as $4\cdot \textit{height} \cdot \textit{width} \cdot \textit{channel}$ in bytes. Before starting the computation of the first fused block $f_1$, the primary ES sends sub-inputs to the secondary ESs. The exchanged data size of $f_1$ is calculated as
\setcounter{equation}{11}
\begin{equation}
	S\left( {{f_1},{\bf{E}}} \right)  = \sum\limits_{{e_k} \in {\bf{E}}\setminus \left\{ {e_1} \right\}} 4\left( \textit{IE}_{f_1}^{e_k} - \textit{IS}_{f_1}^{e_k} + 1 \right) \textit{IF}_{f_1} c_{f_1}^{\textit{in}}
	\label{eq16}
\end{equation}
where $c_{f_1}^{\textit{in}}$ is the number of channels of the input tensor of $f_1$. 

For $f_m, m \in \left[2,M\right]$, ES $e_k$ needs to receive the missing data of its sub-input from its collaborating ESs $e_{k-1}$ and $e_{k+1}$. The received data size between $e_k$ and its collaborating $e_{k-1}$ and $e_{k+1}$, 	${\textit{EX}}_{{f_m}}^{{e_k},{e_{k - 1}}}$ and ${\textit{EX}}_{{f_m}}^{{e_k},{e_{k + 1}}}$, are denoted as (13) and (14). After computing ${f_M}$, the sub-outputs of all the secondary ESs will be sent to the primary ES, which will merge them into the new input for ${f_{M+1}}$. In general, the exchanged data size for $f_m$, $S\left( {{f_m},{\bf{E}}} \right)$, can be calculated as (15), where $c_{f_M}^{\textit{out}}$ is the number of channels of the output tensor of $f_M$. 
\subsection{Minimizing total task inference time}
During the data exchanging phase, we assume that the data transmission is full-duplex and two ESs can communicate with each other simultaneously. With 100 Gbps Ethernet, the communication time for exchanging data is very small compared with the computation time, therefore, the performance of our proposed scheme will be affected marginally, even if it does not use full-duplex transmission mode. Assuming the transmission rate between ESs is $\alpha$, the communication time for exchanging data between ESs for ${f_m}$ can be denoted as
\setcounter{TempEqCnt}{\value{equation}}
\setcounter{equation}{15}
\begin{equation}
	{T^{\textit{com}}}\left( {{f_m},{\bf{E}}} \right) = S\left( {f_m},\bf{E} \right)/\alpha.
	\label{eq19}
\end{equation}

During the computing phase, the primary ES needs to wait until the last collaborating ES completes the computation. Assuming the computation time of ES ${e_k}$ for the fused block ${f_m}$ is $t_{{f_m}}^{{e_k}}$, the computation time for ${f_m}$ can be denoted as
\setcounter{equation}{16}
\begin{equation}
T^{\textit{cmp}}\left( {{f_m},{\bf{E}}} \right)\hspace{-1mm}=\hspace{-1mm} \left\{\hspace{-2mm} {\begin{array}{*{20}{c}}
\max \left( t_{f_m}^{e_1},\cdots, t_{f_m}^{e_K} \right)\!,\hspace{-3mm} &f_m \neq f_{M+1}, \\
t_{{f_m}}^{{e_1}}, & f_m = f_{M+1}.
\end{array}} \right.
\label{eq20}
\end{equation}
Note that $f_{M+1}$ is solely computed at primary ES $e_1$. Hence, the total inference time for ${f_m}$ can be expressed as
\begin{equation}
T^{\textit{inf}}\left( {{f_m},{\bf{E}}} \right) = {T^{\textit{com}}}\left( {{f_m},{\bf{E}}} \right) + {T^{\textit{cmp}}}\left( {{f_m},{\bf{E}}} \right).
\label{eq21}
\end{equation}
The objective to minimize the overall inference time of the inference task is formulated as bellows:
\begin{equation}
\mathop {\min }\limits_{{\bf{E}},{\bf{F}}} \sum\limits_{{f_m} \in {\bf{F}}} {T^{\textit{inf}}\left( {{f_m},{\bf{E}}} \right)},
\label{eq22}
\end{equation}
subject to (\ref{eq5}), (\ref{eq8}) and (\ref{eq9}).
\setcounter{equation}{22}
\begin{figure*}[hb]
	\begin{equation}
		{t^ * }\left( {l_{{f_m}}^{s},l_{{f_m}}^{e}} \right) = \left\{ {\begin{array}{*{20}{c}}
				{t\left( {l_{{f_m}}^{s},l_{{f_m}}^{e}} \right)},&{l_{{f_m}}^{e} = l_{{f_m}}^{s}},\\
				{\mathop {\min }\limits_{1 \le i \le l_{{f_m}}^{e} - l_{{f_m}}^{s} - 1} \left( {t\left( {l_{{f_m}}^{s},l_{{f_m}}^{s} + i} \right) + {t^ * }\left( {l_{{f_m}}^{s} + i + 1,l_{{f_m}}^{e}} \right)} \right)},&{l_{{f_m}}^{s} < l_{{f_m}}^{e} \le N}.
		\end{array}} \right.
	\end{equation}
	\vspace{-6mm}
\end{figure*}
\begin{algorithm}
\caption{Dynamic Programming for fused-layer parallelization (DPFP)}
\KwIn{CNN model ${\bf{G}}$, selected ESs ${\bf{E'}}$, Optimal CNN model partitioning point set ${\bf{S}}$ }
\KwOut{Optimal Fused blocks ${{\bf{F}}^*}$}
\textbf{Step 1:} Initialize ${t^ * }\left(  \cdot  \right) \leftarrow MAX$, ${\bf{S}} \leftarrow \emptyset $ \\
\textbf{Step 2:} Define a function ${\mathop{\rm \textbf{DPFP}}\nolimits} \left({i,j} \right)$:\\
\textbf{function ${\mathop{\rm \textbf{DPFP}}\nolimits} \left({i,j} \right)$:}\\
  \eIf{$j = i$}{
     return ${t^ * }\left( {i,j} \right)\leftarrow t\left( {i,j} \right)$ and ${\bf{S}} \leftarrow {\bf{S}} \cup j$ \;
  }{
     let $t_{\textit{min}} = MAX$ \\
    \For{$k$ range in $\left[ {1,j - i-1} \right]$}{
        $t = t\left( {i,i + k} \right) + {\mathop{\rm DPFP}\nolimits} \left( {i + k + 1,j} \right)$ \\
     \If{$t < t_{\textit{min}}$}{
      $t_{\textit{min}} = t$ and the optimal partitioning point $l_{\textit{opt}} = i + k$ \;
     }
    }
    return ${t^ * }\left( {i,j} \right)\leftarrow  t_{\textit{min}}$ and ${\bf{S}} \leftarrow {\bf{S}} \cup l_{\textit{opt}}$\;
  }
\textbf{Step 3:} Using step 2 to calculate ${t^ * }\left( {1,N} \right)$ and ${\bf{S}}$, then ${{\bf{F}}^*}$ can be computed according to ${\bf{S}}$.
\end{algorithm}
\section{Dynamic Programming for fused-layer parallelization}\label{solution}
The optimization goal of (19) is to search the optimal number of ESs and the optimal partition of a CNN model. Therefore, this minimization problem can be divided into the two steps: (i) to determine optimal number of the fused blocks of the CNN model for the given number of ESs; (ii) to determine the optimal number of ESs. Assuming the selected ESs ${{\bf{E}}'}$ and the ratio $\eta _{{f_m}}^{{e_k}}$ are given, ${T^{\it{com}}}\left( {{f_{M+1}},{\bf{E}}'} \right)$ and ${T^{\it{cmp}}}\left( {{f_{M+1}},{\bf{E}}'} \right)$ are determined no matter how to partition the CNN model. Therefore, the minimization problem (\ref{eq22}) is equivalent to minimize the completion time of all the CLs as
\setcounter{TempEqCnt}{\value{equation}}
\setcounter{equation}{19}
\begin{equation}
	\mathop {\min }\limits_{{\bf{F}}} \sum\limits_{{f_m} \in {\bf{F}}\setminus{f_{M+1}}} {T^{\textit{inf}}\left( {{f_m},{\bf{E}}'} \right)}.
\label{eq21}
\end{equation}
For ease of description, the inference time $T^{\textit{inf}}\left( {{f_m},{\bf{E}}'} \right)$ of the fused block ${f_m},m\neq M+1 $ can be simplified as $t\left( l_{f_m}^{s},l_{f_m}^{e} \right)$. Assuming the optimal partitioning of ${\bf{G}}$ is ${{\bf{F}}^ * }$ with ${M^ *+1 }$ fused blocks, the minimal inference time of all the CLs in (\ref{eq21}) is equal to the sum of the inference time of each fused block, which can be denoted as
\begin{eqnarray}
	t^{*}\left(1,N\right)=\sum_{f_m \in {{\bf{F}}^*}\setminus{f_{M^*+1}} } t\left( l_{f_m}^s,l_{f_m}^e \right).
	\label{eq24}
\end{eqnarray}
This optimization problem can be solved by the well-known rod-cutting problem in dynamic programming \cite{CLRS2009dynamicprograming}. More generally, the optimal inference time can be expressed as
\begin{eqnarray} 
{t^ * }\left( {1,N} \right)\hspace{-1mm} & = &\hspace{-1mm} \min \left\{ {t(1,N),{t^ * }\left( {1,2} \right) + {t^ * }\left( {3,N} \right),{t^ * }\left( {1,3} \right) + } \right. \nonumber\\
& &\left. {{t^ * }\left( {4,N} \right), \cdots ,{t^ * }\left( {1,N - 1} \right) + t\left( {N,N} \right)} \right\}\hspace{-1mm}.
\label{eq23}
\end{eqnarray}

Simplifying (\ref{eq23}), the minimal inference time for ${f_m}$ can be expressed as (23), below. Based on dynamic programming, the optimal set for the fused blocks can be obtained as described in Algorithm 1. After the optimal fused blocks ${{\bf{F}}^ * }$ are determined for the selected ESs ${\bf{E}}'$, the next step is to derive the optimal ESs by comparing the inference time under different number of the selected ESs. The one that has the lowest inference time is the optimal selection of ESs ${\bf{E}}'$. To compare the performance of the proposed collaborative edge computing for distributed CNN with the conventional approach, we define the speedup ratio $\rho $ for the optimal ${\bf{E}}'$ as
\setcounter{TempEqCnt}{\value{equation}}
\setcounter{equation}{23}
\begin{equation}
\rho = 1-T^{\textit{inf}}\left( \bf{E}' \right)/T^{\it{pre}}.
\label{eq25}
\end{equation}
where $T^{\textit{inf}}\left( {{\bf{E}}'} \right)$ is the inference time for ESs ${\bf{E}}'$, ${T^{\it{pre}}}$ is the inference time of the pre-trained model running on standalone ES. It is clear that the faster the proposed approach is, the higher speedup ratio will be achieved.
\begin{table}[h]
 \begin{minipage}[t]{0.5\textwidth}
 	\vspace{-1mm}
	\centering
	\caption{Inference time for variable ESs at 100 G{\upshape bps} between ESs({\upshape ms})}
	\label{tab:3}
	\vspace{-2mm}
	\scalebox{0.9}{
	\begin{tabular}{|c|c|c|c|c|c|c|c|c|c|c|c|c|c|}
		\hline
		\multirow{2}*{} & \multicolumn{3}{c|}{2 ESs}  & \multicolumn{3}{c|}{7 ESs}\\
		\cline{2-7}
		& \multicolumn{1}{c|}{} & \multicolumn{1}{c|}{} & \multicolumn{1}{c|}{} & \multicolumn{1}{c|}{} & \multicolumn{1}{c|}{}& \multicolumn{1}{c|}{}\\[-0.9em]
		& {$T^{\textit{cmp}}$} & {$T^{\textit{com}}$} & $T^{\textit{inf}}$ &   {$T^{\textit{cmp}}$} & {$T^{\textit{com}}$} & $T^{\textit{inf}}$     \\ \hline
	    {\bf DPFP\_RTX 2080TI} & 2.26 & 0.08 &2.34  & 1.32 & 0.35 & 1.67  \\
	    {\bf DPFP\_GTX 1080TI} & 2.79 &0.08 & 2.87  & 1.53  & 0.35 &1.88 \\
  	    {\bf DPFP\_AGX Xavier}  & 16.69  &0.10 &16.79  & 8.20 & 0.52 &8.72\\        \hline
	\end{tabular}
	}
\end{minipage}
\begin{minipage}[t]{0.5\textwidth}
\centering
	\caption{Inference time for 7 ESs with variable transmission rate ({\upshape ms})}
	\label{tab:4}
	\vspace{-2mm}
	\scalebox{0.85}{
	\begin{tabular}{|c|c|c|c|c|c|c|c|}
		\hline
		\multirow{2}*{} & \multicolumn{3}{c|}{40Gbps} & \multicolumn{3}{c|}{100Gbps} \\
		\cline{2-7} & \multicolumn{1}{c|}{} & \multicolumn{1}{c|}{} & \multicolumn{1}{c|}{} & \multicolumn{1}{c|}{} & \multicolumn{1}{c|}{}& \multicolumn{1}{c|}{}\\[-0.9em]
		& {$T^{\it{cmp}}$} & {$T^{\it{com}}$} & $T^{\textit{inf}}$ & {$T^{\it{cmp}}$} & {$T^{\it{com}}$} & $T^{\textit{inf}}$ \\
		\hline
		{\bf DPFP\_RTX 2080TI}  & 1.38 & 0.74 & 2.12 & 1.32 & 0.35 & 1.67\\
		{\bf DPFP\_GTX 1080TI}   & 1.59 & 0.74 & 2.33 & 1.53 & 0.35 & 1.88\\
		{\bf DPFP\_AGX Xavier}  & 8.24 & 1.30 & 9.54 & 8.20 & 0.52 & 8.72\\
		{ MoDNN\_RTX 2080TI} & 1.01 & 9.97 & 10.98 & 1.03 & 3.98 & 5.01\\
		{ MoDNN\_GTX 1080TI}  & 1.12 & 9.97 & 11.09 & 1.15  & 3.98 & 5.13\\
		{ MoDNN\_AGX Xavier}  &8.23 & 9.97 & 17.20 & 8.22 & 3.98 & 11.20 \\
		\hline
	\end{tabular}}
\end{minipage}
\vspace{-4mm}
\end{table}
\begin{figure*}[htbp]
\centering
\begin{minipage}[t]{0.48\textwidth}
\centering
\includegraphics[width=0.6\textwidth]{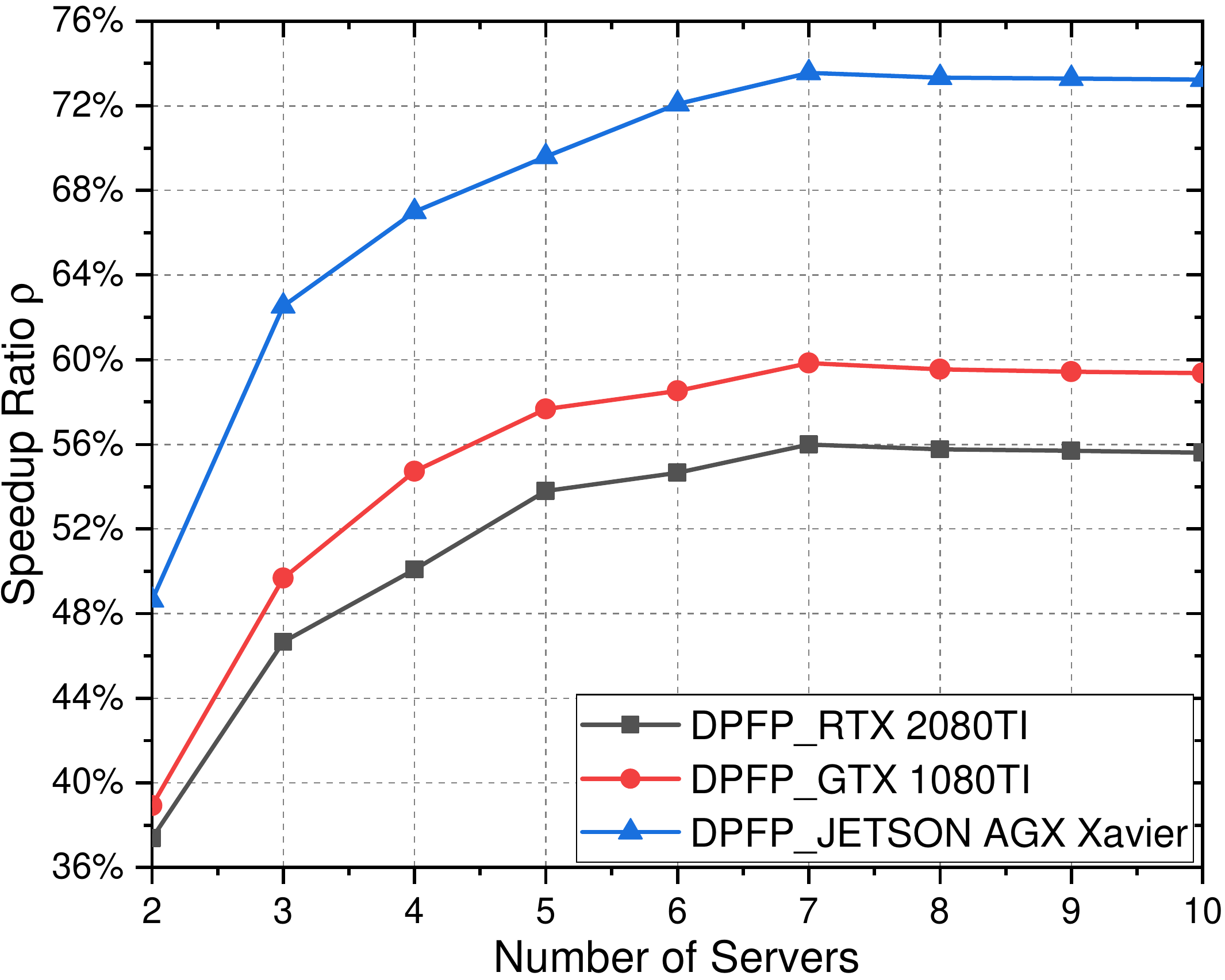}
\vspace{-2mm}
\caption{Speedup ratio vs. No. of ESs at 100Gbps between ESs.}
\label{Fig.4}
\end{minipage}
\begin{minipage}[t]{0.48\textwidth}
\centering
\includegraphics[width=0.6\textwidth]{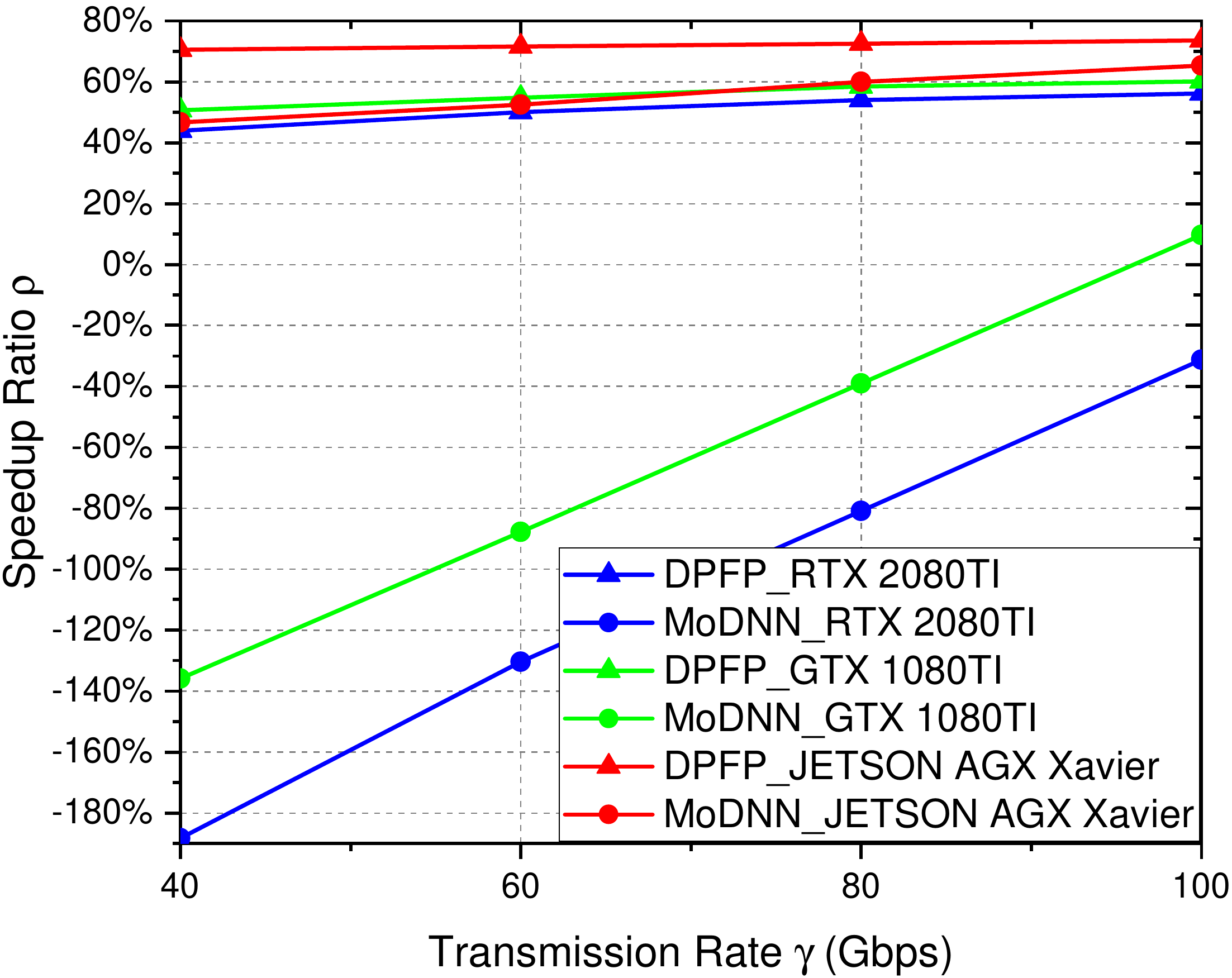}
\vspace{-2mm}
\caption{Speedup ratio vs. variable transmission rate when 7 ESs collaborate.}
\label{Fig.5}
\end{minipage}
\vspace{-5mm}
\end{figure*}
\addtolength{\topmargin}{0.06in}
\section{Performance Evaluation}\label{simulation}
\subsection{Simulation Setup and Methodology}
In this simulation, the edge computing network consists of one IoT device and 10 ESs are of equal computing capacity. The data communication between ESs via Ethernet with transmission rate $\alpha$ ranging from 40Gbps to 100Gbps \cite{032}. We use the image recognition model VGG-16 \cite{004} in the experiments. We compare the proposed approach with a state-of-the-art work MoDNN \cite{005}. To provide a realistic performance evaluation, we measured computation time on three different GPU platforms, i.e., RTX 2080TI with 13.45 TFLOPS of float32, GTX 1080TI with 11.3 TFLOPS of float32, and JETSON AGX Xavier with 1.41 TFLOPS of float32. The shape of the input images is $224 \times 224 \times 3$, and the performance metric is speedup ratio $\rho $.
\subsection{Impact of ESs}\label{No.ES}
TABLE \ref{tab:3} shows the computation time and communication time with different number of ESs at the transmission rate equal to 100Gbps. Generally, it can be seen that as the number of ESs increases, the computation time decreases, on the contrary, the communication time between ESs increases. This is because when more ESs are sharing the inference task, it will naturally reduce the sub-task size on each ES; however, the total exchanged data size between ESs will increase. When the number of ESs is small, the computation time takes up most of the overall inference time; therefore, the speedup ratio will increase as the number of ESs increases, as shown in Fig. \ref{Fig.4}. When the number of ESs is larger than 7, the reduced computation time by using more ESs is rather small and is even close to the introduced extra communication time. Therefore, the speedup ratio reaches a plateau. 

Comparing the performance of different platforms, it can be seen that, the optimal sets of the fused blocks for GTX 1080TI and RTX 2080TI are almost the same, however, for JETSON AGX Xavier the model is in fact partitioned at every CL. This is because to strike a good balance between the communication time and computation time, it is worth fusing several CLs into one fused block when an ES has high computation capacity, e.g., for the cases of GTX 1080TI and RTX 2080TI; however, for low computation capacity platform, e.g., JETSON AGX Xavier, it does not pay off to fuse several CLs into one fused block. In other words, the extra computation time caused by fused block is more than the saved communication time for low computation capacity platforms.
\subsection{Impact of transmission rate between ESs}
TABLE \ref{tab:4} shows the inference time at different transmission rates between ESs for the case of 7 ESs collaborating.
We can see DPFP partitions the model at each CL for JETSON AGX Xavier as MoDNN, whereas for the other two platforms DPFP fuses several CLs into one block. Since computation capacities of the RTX 2080TI and GTX 1080 TI do not have a huge difference, the optimal partitioning solution is the same for these two platforms. Moreover, the partitioning solution varies under different transmission rates in the case of RTX 2080TI and GTX 1080TI. For example, we can observe that the exchanged data increases from 3.7 MBytes at 40 Gbps to 4.375 MBytes at 100 Gbps, which indicates DPFP partitions the model into smaller blocks at a higher transmission rate, to trade off the computation time and communication time. As MoDNN needs to exchange data for merging sub-outputs after every CL and DPFP only exchanges the overlapped data between ESs after each fused block, DPFP has much less communication overhead and can reduce the communication time about 90\% at a given transmission rate.

Therefore, as shown in Fig. \ref{Fig.5}, using the same platform DPFP has better speedup ratio than MoDNN. As the transmission rate increases, the speedup ratio of a platform using DPFP will improve and then reach a plateau. This is because with increasing transmission rate, the communication time gradually decreases and takes up less and less time of the overall inference time, while the computation time nearly keeps unchanged. In addition, the difference of speed up ratio of DPFP and MoDNN depends on the platform's computing capacity for a given transmission rate. The higher computation capacity is, the higher gain can be achieved. This is because the platform with higher computing capacity uses less computation time and the communication time is a big part of the overall inference time.
\begin{table*}[]
\centering
\caption{Service reliability on RTX 2080TI under different time-variant channels (over 99.999\% in bold)}
\vspace{-2mm}
\label{tab:7}
\scalebox{0.85}{
\begin{tabular}{|c|c|c|c|c|c|c|c|}
\hline
\multirow{2}{*}{\begin{tabular}[c]{@{}c@{}}Number \\ of ESs\end{tabular}}
                                 & \multicolumn{2}{c|}{40 Mbps}   & \multicolumn{2}{c|}{60 Mbps} & \multicolumn{3}{c|}{100 Mbps}          \\ \cline{2-8} 
& \begin{tabular}[c]{@{}c@{}}$\delta=1$ ms \\ ($\phi=4.3$ Mbps) \end{tabular}
& \begin{tabular}[c]{@{}c@{}}$\delta=2$ ms \\ ($\phi=7.7$ Mbps)\end{tabular}     
& \begin{tabular}[c]{@{}c@{}}$\delta=2$ ms \\ ($\phi=15.9$ Mbps)\end{tabular}        
& \begin{tabular}[c]{@{}c@{}}$\delta=3$ ms \\ ($\phi=21.2$ Mbps)\end{tabular}      
& \begin{tabular}[c]{@{}c@{}}$\delta=3$ ms \\ ($\phi=47.4$ Mbps)\end{tabular}    
& \begin{tabular}[c]{@{}c@{}}$\delta=4$ ms \\ ($\phi=54.5$ Mbps)\end{tabular}     
& \begin{tabular}[c]{@{}c@{}}$\delta=5$ ms \\ ($\phi=60.0$ Mbps)\end{tabular}               \\ \hline
1                                                                         & 0.945201 & 0.788145 & {\textbf{1}} & 0.997168     &{\textbf{1}}  & 0.999983       & 0.999550            \\ \hline
2                                                                         & {\textbf{0.999996}} & 0.986791 & {\textbf{1}} & 0.999402      & {\textbf{1}} &{\textbf{0.999999}}    & 0.999959         \\ \hline
6                                                                         & {\textbf{1}}        & 0.997154 & {\textbf{1}}  & 0.999689      &{\textbf{1}} &{\textbf{1}}    & {\textbf{0.999990}}         \\ \hline
\end{tabular}}
\vspace{-5mm}
\end{table*}
\subsection{Evaluation of service reliability}
Instead of offloading the sub-tasks to ESs directly from IoT device, the primary ES makes decision on how to partition the inference task and how to distribute the sub-tasks to the secondary ESs. This full offloading scheme simplifies the processing and minimizes resource demands on IoT device. For an inference task, the total task completion time can be denoted as $T = T^{\textit{off}} + T^{\textit{inf}}$, where $T^{\textit{off}}$ is offloading time of an inference task transmitted from IoT device to primary ES.

To ensure real-time inference, we can decide the minimal transmission data rate between IoT device and the primary ES according to Section \ref{No.ES}. For example, to ensure system throughput (e.g., 30 FPS) for an input image (e.g., 125 KBytes), the transmission data rate between IoT device and the primary ES should not be lower than 32 Mbps. Generally, the channel state is stochastic, which may cause fluctuations in offloading time $T^{\textit{off}}$. Assume the offloading time $T^{\textit{off}}\sim\mathcal{N}(\mu,\,\delta^{2})$, we can estimate the fluctuation of transmission rate, $\phi$, based on the three-sigma rule of thumb. In mission critical IoT services, the required service reliability (the probability that inference feedback meets the service deadline) can be in the range of 99\% to 99.999\%. 

TABLE \ref{tab:7} shows the service reliability under the time-variant channel of different transmission rate. It can be seen that the service reliability will decrease as the fluctuation of channel status increases. To address this issue, a higher transmission rate can contribute to improving the service reliability. However, relying solely on improving transmission rate cannot guarantee the high service reliability of 99.999\%. For example, a standalone ES can achieve max 99.955\% service reliability for the scenario at average data rate of 100 Mbps and fluctuation $\phi= 60.0$ Mbps. Note that in a mobile network, large fluctuation in data rate is likely to occur, for example, the uplink data rate of a moving vehicle. In this case, DPFP can still ensure high service reliability of 99.999\% and even further.

\section{Conclusion}\label{conclusion}
This paper studies inference acceleration through distributed CNN using collaborative edge computing for time-critical IoT applications. We propose RFS to effectively partition an inference task without compromising the inference accuracy of the pre-trained model. To minimize the communication and computing costs, we propose a novel fused-layer parallelization using collaborative edge computing to partition a CNN model into fused blocks. In addition, we use dynamic programming to effectively solve the optimization problem of minimizing total inference time. The experimental results show that DPFP can accelerate inference of VGG-16 up to 73\% compared with the pre-trained model, which outperforms the state-of-the art approach MoDNN. Furthermore, we evaluate the service reliability of DPFP under time-variant channel, which shows that DPFP is an effective solution to ensure high service reliability with strict service deadline.
\ifCLASSOPTIONcaptionsoff
  \newpage
\fi

\section*{Acknowledgment}
This work is supported by Agile-IoT project (Grant No. 9131-00119B) granted by the Danish Council for Independent Research.

\bibliographystyle{IEEEtran}
\bibliography{ddnn}

\end{document}